\documentclass{llncs}
\makeatletter
\renewcommand\subsubsection{\@startsection{subsubsection}{3}{\z@}%
  {-18\p@ \@plus -4\p@ \@minus -4\p@}%
  {8\p@ \@plus 4\p@ \@minus 4\p@}%
  {\normalfont\normalsize\bfseries\boldmath\rightskip=\z@
    \@plus 8em \pretolerance=10000 }}
\makeatother

\usepackage{etex,amsfonts,amsmath,ulem}
\usepackage{mathbbol,graphics,epsf,array}
\usepackage{verbatim,amssymb,amscd,eucal,bbm,setspace}
\usepackage[np]{numprint}
\usepackage[tight,scriptsize]{subfigure}
\usepackage{listings,gag}
\lstdefinelanguage{GAP}[]{Pascal}{%
    morekeywords={od,fi,alias,local,return,fail},%
    deletekeywords={with},
    keywordstyle=\underline, %
    morecomment=[l]{\#},
    commentstyle=\mdseries, %
    breaklines=false, %
    basicstyle=\ttfamily\small, %
    sensitive=false,
    showspaces=false, %
    showlines=true, %
    showstringspaces=false, %
    backgroundcolor=\color[gray]{0.95}, %
    breakautoindent=true, %
    tabsize=4, %
}
\usepackage{graphicx}
\usepackage{vaucanson-g}
\usepackage{array,multirow,colortbl}%
\usepackage{rotating}
\usepackage{enumerate,textcomp}
\usepackage{tikz}
\makeatletter
\renewenvironment{table}
  {\setlength\abovecaptionskip{-10\p@}%
   \setlength\belowcaptionskip{0\p@}%
   \setlength\intextsep{20\p@}%
   \@float{table}}
  {\end@float}
\renewenvironment{figure}
  {\setlength\abovecaptionskip{0\p@}%
   \setlength\belowcaptionskip{-10\p@}%
   \@float{figure}}
  {\end@float}
\makeatother

\renewcommand{\thesubfigure}{\thefigure.\arabic{subfigure}} \makeatletter
\renewcommand{\p@subfigure}{}
\renewcommand{\@thesubfigure}{\thesubfigure:\hskip\subfiglabelskip} \makeatother

\usepackage{enumitem}
\setlist{nolistsep}

\newcolumntype{E}{!{\vrule width 1.3pt}}

\definecolor{verylightgray}{gray}{0.95}

\newcommand{\pres}[1]{\langle{#1}\rangle}
\newcommand{\presm}[1]{\pres{{#1}}_{+}}

\def\FR{{\sf FR}}
\def\SK{{\sf automgrp}}
\def\GAP{{\sf GAP}}
\def\resp{\hbox{\textit{resp.}} }

\def\N{\mathbb{N}}

\newcounter{foo}

\def\mz{{\mathfrak m}}
\def\dz{{\mathfrak d}}
\newcommand{\Ac}{\mathcal{A}}

\newcommand{\mdreduc}[1]{\mz\dz^*{(#1)}}

\newcommand{\tensorpower}[2]{{#1}_{#2}}

\newcommand{\ordreElement}[1]{}

\newcolumntype{E}{!{\vrule width 1.3pt}}

\title{Implementing Computations\\
in Automaton (Semi)groups}
\author{Ines Klimann \and Jean Mairesse \and Matthieu Picantin}
\institute{Univ Paris Diderot, Sorbonne Paris Cit\'e,
 LIAFA, UMR 7089 CNRS, Paris, France\\
\email{\{klimann,mairesse,picantin\}@liafa.univ-paris-diderot.fr}}
\date{\today}

\begin{document}

\maketitle

\begin{abstract}
We consider the growth, order, and finiteness problems for automaton
(semi)groups. We propose new implementations 
and compare them with the existing ones. As a result of extensive
experimentations, we propose some
conjectures on the order of finite automaton (semi)groups.
\end{abstract}


\normalem

\section{Introduction}\label{s:intro}

\bigskip\emph{Automaton (semi)groups} --- short for semigroups generated by
Mealy automata or groups generated by invertible Mealy automata ---
were formally introduced a half century ago (for details,
see~\cite{gns,cain} and references therein).
Over the years, important results have started revealing their full
potential.
For instance, the article~\cite{grigorchuk1} constructs simple
Mealy automata generating infinite torsion groups and so contributes
to the Burnside problem, and, the article~\cite{brs} produces Mealy
automata generating the first examples of (semi)groups with
intermediate growth and so answers the Milnor problem.

The classical decision problems have been investigated for such (semi)groups. The 
word problem is solvable using standard minimization
techniques, while the conjugacy problem is undecidable~\cite{conjugacy}.
Here we concentrate on the problems related to growth, order, and finiteness.

\begin{figure}[h]
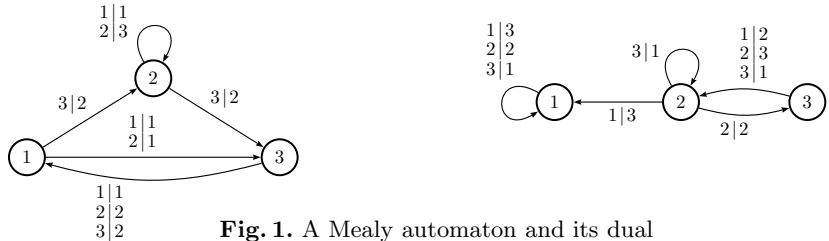

\centering
\subfigure{%
\label{M33dual}
\TinyPicture
\VCDraw{%
\begin{VCPicture}{(-3,-1)(5,4)}
\LargeState
\State[1]{(-4,0.5)}{a}
\State[3]{(4,0.5)}{b}
\State[2]{(0,3)}{c}
\EdgeL{a}{c}{\IOL{3}{2}}
\EdgeL{c}{b}{\IOL{3}{2}}
\LoopN{c}{\StackTwoLabels{\IOL{1}{1}}{\IOL{2}{3}}}
\EdgeL{a}{b}{\StackTwoLabels{\IOL{1}{1}}{\IOL{2}{1}}}
\ArcL[.7]{b}{a}{\StackThreeLabels{\IOL{1}{1}}{\IOL{2}{2}}{\IOL{3}{2}}}
\end{VCPicture}}
}\hspace*{3cm}
\subfigure{%
\label{M33}
\TinyPicture
\VCDraw{%
\begin{VCPicture}{(-4,-2.5)(4.5,1)}
\LargeState
\State[1]{(-4,0)}{a}
\State[2]{(0,0)}{b}
\State[3]{(4,0)}{c}
\LoopW{a}{\StackThreeLabels{\IOL{1}{3}}{\IOL{2}{2}}{\IOL{3}{1}}}
\LoopN{b}{\IOL{3}{1}}
\EdgeL{b}{a}{\IOL{1}{3}}
\ArcR{b}{c}{\IOL{2}{2}}
\ArcR{c}{b}{\StackThreeLabels{\IOL{1}{2}}{\IOL{2}{3}}{\IOL{3}{1}}}
\end{VCPicture}}
}
\caption{A Mealy automaton and its dual}
\label{monster}
\end{figure}

To illustrate, consider the two Mealy automata of Fig.~\ref{monster}. They
are dual, that is, they can be obtained one from the other by
exchanging the roles of stateset and alphabet. A (semi)group is
associated in a natural way with each
  automaton (formally defined below). The two Mealy automata of
Fig.~\ref{monster} are associated with finite (semi)groups. Their
orders are respectively: on the left a semigroup of order~238, on
the right a group of order~\(\numprint{1494186269970473680896}=2^{64}\cdot 3^4\approx
1.5\times 10^{21}\).

\bigskip

Several points are illustrated by this example:
\begin{itemize}
\item An automaton and its dual generate (semi)groups which are either both finite or both infinite (see~\cite{Nek,AKLMP12}).
\item The order of a finite automaton (semi)group can be amazingly
  large. It makes a priori difficult to decide whether an
  automaton (semi)group is finite or not. Actually, the decidability of
  this question is open (see~\cite{gns,AKLMP12}).
\item The order of the (semi)groups generated by a Mealy automaton and its dual can be strikingly different. It suggests to work with both automata together.
\end{itemize}

\smallskip

The contributions of the present paper are three-fold: 
\begin{itemize}
\item We propose new implementations (in \GAP~\cite{GAP4}) of
  classical algorithms for the computation of the growth
  function; the computation of the order (if finite); the semidecision
  procedure for the finiteness.
\item We compare the new implementations with the existing ones. Indeed, there exist two \GAP\ packages dedicated to Mealy automata and their
associated (semi)groups: \FR\ by~Bartholdi~\cite{FR} and~\SK\ by~Muntyan
and~Savchuk~\cite{sav}. 
\item We realize systematic experimentations on small Mealy automata
  as well as randomly chosen large Mealy automata. These serve as
  testbeds to some conjectures on the growth types of the associated
  (semi)groups, as well as on the order of \ordreElement{an element or
  }a (semi)group. 
\end{itemize}

\smallskip

The structure of the paper is the following. In
Section~\ref{sec:semigroups}, we present basic notions on Mealy automata
and automaton (semi)groups. In Section~\ref{sec:minimizing}, we
give new implementations and compare them with the existing
ones. Section~\ref{sec:conjectures} is dedicated to experimentations
and to the resulting conjectures.

\section{Automaton (Semi)groups}\label{sec:semigroups}

\subsection{Mealy Automaton}
If one forgets initial and final states, a {\em
(finite, deterministic, and complete) automaton} $\aut{A}$ is a
triple 
\(
\bigl( A,\Sigma,\delta = (\delta_i: A\rightarrow A )_{i\in \Sigma} \bigr)
\),
where the \emph{set of states}~$A$
and the \emph{alphabet}~$\Sigma$ are non-empty finite sets, and
where the $\delta_i$'s are functions.

\smallskip

A \emph{Mealy automaton} is a quadruple 
\[
\bigl( A, \Sigma, \delta = (\delta_i: A\rightarrow A )_{i\in \Sigma},
\rho = (\rho_x: \Sigma\rightarrow \Sigma  )_{x\in A} \bigr) \:,
\]
such that both $(A,\Sigma,\delta)$ and $(\Sigma,A,\rho)$ are
automata. 
In other terms, a Mealy automaton is a letter-to-letter transducer
with the same input and output alphabets. 
The transitions of a Mealy automaton are
\[x\xrightarrow{i | \rho_x(i)} \delta_i(x) \:.\]

The graphical representation of a Mealy automaton is standard, see
Fig.~\ref{monster}.

The notation $x\xrightarrow{\mot{u}| \mot{v}} y$ with
$\mot{u}=u_1\cdots u_n$, $\mot{v}=v_1\cdots v_n$ is a shorthand for
the existence of a path $x \xrightarrow{u_1|v_1} x_1
\xrightarrow{u_2|v_2} x_2 \longrightarrow \cdots \longrightarrow x_{n-1}
\xrightarrow{u_{n}|v_n} y$ in $\aut{A}$. 

\smallskip

In a Mealy automaton $(A,\Sigma, \delta, \rho)$, the sets $A$ and
$\Sigma$ play dual roles. So we may consider the \emph{dual (Mealy)
  automaton} defined by
\(
\dz(\aut{A}) = (\Sigma,A, \rho, \delta)
\), that is:
\begin{equation*}
i \xrightarrow{x\mid y} j \ \in \dz(\aut{A}) \quad \iff \quad x
\xrightarrow{i\mid j} y \ \in \aut{A} \:.
\label{eq-dual}
\end{equation*}
It is pertinent to consider a Mealy automaton
and its dual together, that is to work with the pair $\{\aut{A},
\dz(\aut{A})\}$, 
see an example in Fig.~\ref{monster}.

\medskip
Let \(\aut{A}=(A,\Sigma,\delta,\rho)\)
and~\(\aut{B}=(B,\Sigma,\gamma,\pi)\) be two Mealy automata acting on
the same alphabet; 
their \emph{product}~\(\aut{A}\times\aut{B}\) is defined as the Mealy
automaton with stateset \(A\times B\), 
alphabet \(\Sigma\), and transitions:
\[xy\xrightarrow{i | \pi_y(\rho_x(i))}
\delta_i(x)\gamma_{\rho_x(i)}(y)\:.\]

\subsection{Generating (Semi)groups}\label{sse-automgroup}
Let $\aut{A} = (A,\Sigma, \delta,\rho)$ be a Mealy automaton. 
We view $\aut{A}$ as an automaton with an input and an output tape, thus
defining mappings from input words over~$\Sigma$ to output words
over~$\Sigma$. 
Formally, for $x\in A$, the map
$\rho_x : \Sigma^* \rightarrow \Sigma^*$,
extending~$\rho_x : \Sigma \rightarrow \Sigma$, is defined by:
\[
\rho_x (\mot{u}) = \mot{v} \quad \textrm{if} \quad \exists y,
\ x\xrightarrow{\mot{u}|\mot{v}} y \:.\]
By convention, the image of the empty word is itself.
The mapping~$\rho_x$ is length-preserving and prefix-preserving.
It satisfies 
\begin{equation*}
\forall u \in \Sigma, \ \forall \mot{v} \in \Sigma^*, \qquad
\rho_x(u\mot{v}) = \rho_x(u)\rho_{\delta_u(x)}(\mot{v}) \:.
\end{equation*} 
We say that $\rho_x$ is the \emph{production
function} associated with
$(\aut{A},x)$. 
For~$\mot{x}=x_1\cdots x_n \in A^n$ with~$n>0$, set
\(\rho_\mot{x}: \Sigma^* \rightarrow \Sigma^*, \rho_\mot{x} = \rho_{x_n}
\circ \cdots \circ \rho_{x_1} \:\).

Denote dually by $\delta_i:A^*\rightarrow A^*,
i\in \Sigma$, the production mappings associated with
the dual Mealy automaton~$\dz(\aut{A})$. For~$\mot{v}=v_1\cdots v_n
\in \Sigma^n$ with~$n>0$, set $\delta_\mot{v}: A^* \rightarrow A^*,
\ \delta_\mot{v} = \delta_{v_n}\circ \cdots \circ \delta_{v_1}$.

\begin{definition}
Consider a Mealy automaton $\aut{A}$.
The semigroup of mappings from~$\Sigma^*$ to~$\Sigma^*$ generated by
$\rho_x, x\in A$, is called the \emph{semigroup generated
  by~$\aut{A}$} and is denoted by~$\presm{\aut{A}}$.
A semigroup $G$ is an \emph{automaton semigroup} if there exists
a Mealy automaton $\aut{A}$ such that $G = \presm{\aut{A}}$.
\end{definition}

A Mealy automaton \(\aut{A}=(A,\Sigma,\delta,\rho)\) is
\emph{invertible} if all the mappings \(\rho_x:\Sigma\to\Sigma\) are 
  permutations. Then the production functions
  \(\rho_x:\Sigma^*\to\Sigma^*\) are invertible.

\begin{definition}
Let \(\aut{A}=(A,\Sigma,\delta,\rho)\) be invertible. The \emph{group
  generated by \(\aut{A}\)} is the group generated by the mappings
  \(\rho_x:\Sigma^*\to\Sigma^*\), \( x\in A\). It is denoted
by~\(\pres{\aut{A}}\).
\end{definition}

Let \(\aut{A}=(A,\Sigma,\delta,\rho)\) be an invertible Mealy
automaton. Its \emph{inverse} is the Mealy automaton \(\inverse{\aut{A}}\)
with stateset \(A^{-1}=\{x^{-1},x\in A\}\) and set of transitions
\begin{equation*}
x^{-1} \xrightarrow{j\mid i} y^{-1} \ \in \aut{A}^{-1} \quad \iff
\quad x \xrightarrow{i\mid j} y \ \in \aut{A} \:.
\end{equation*}

A Mealy automaton is \emph{reversible} if its dual is invertible.
A Mealy automaton \(\aut{A}\) is \emph{bireversible} if both
\(\aut{A}\) and \(\inverse{\aut{A}}\) are invertible and reversible.

\begin{theorem}[\cite{AKLMP12,Nek,sv11}]\label{finitedualfinite}
The (semi)group generated by a Mealy automaton is finite
if and only if the (semi)group generated by its dual is finite.
\end{theorem}

\subsection{Minimization and the Word Problem}

Let $\Ac=(A,\Sigma,\delta,\rho)$ be a Mealy automaton.
The \emph{Nerode equivalence on \(A\)} is the limit of the sequence of
increasingly finer equivalences~$(\equiv_k)$ recursively defined by:
\begin{align*}
\forall x,y\in A,\qquad\qquad x\equiv_0 y & \ \Longleftrightarrow \ \rho_x=\rho_y\:,\\
\forall k\geqslant 0,\, x\equiv_{k+1} y & \ \Longleftrightarrow\  x\equiv_k y\quad
 \text{and}\quad\forall i\in\Sigma,\ \delta_i(x)\equiv_k\delta_i(y)\:.
\end{align*}
Since the set $A$ is finite, this sequence is ultimately constant; moreover if
two consecutive equivalences are equal, the sequence remains constant from this
point. The limit is therefore computable.
For every element~$x$ in~$A$, we denote by~$[x]$ the class of~$x$ w.r.t. the Nerode equivalence.

\begin{definition}
Let $\Ac=(A,\Sigma,\delta,\rho)$ be a Mealy automaton
and let $\equiv$ be the Nerode equivalence on \(A\).
The \emph{minimization} of $\Ac$ is the Mealy automaton
\(\mz(\Ac)=(A/\mathord{\equiv},\Sigma,\tilde{\delta},\tilde{\rho})\),
where
for every $(x,i)$ in $A\times \Sigma$,
$\tilde{\delta}_i([x])=[\delta_i(x)]$ and
$\tilde{\rho}_{[x]}=\rho_x$.
\end{definition}

This definition is consistent with the standard minimization of ``deterministic
finite automata'' where instead of
considering the mappings $(\rho_x:\Sigma\to\Sigma)_x$, the computation
is initiated by the separation between terminal and non-terminal
states. Using Hopcroft algorithm, the time complexity of minization is
\({\cal O}(\Sigma A\log{A})\), see~\cite{ahu74}.

\smallskip

By construction,
a Mealy automaton and its minimization generate the same
semigroup.
Indeed, 
two states of a Mealy automaton belong to the same class w.r.t the
Nerode equivalence if and only if they represent the same element in
the generated (semi)group.

\smallskip

Consider the \emph{word problem}:

\smallskip

\qquad
\begin{minipage}{.9\linewidth}
{\bf Input:} a Mealy automaton \((A,\Sigma,\delta,\rho)\);
\(\mot{x},\mot{y}\in A^*\). 

{\bf Question:} \((\rho_{\mot{x}}:\Sigma^*\to\Sigma^*)=
(\rho_{\mot{y}}:\Sigma^*\to\Sigma^*)\)?
\end{minipage}
\ \\

The word problem is solvable by extending the above minimization
procedure. \FR\ uses this approach, while \SK\ uses a method based on
the wreath recursion~\cite{cain}.

\section{Fully Exploiting the Minimization}\label{sec:minimizing}
Consider the following problems for the (semi)group given by a
Mealy automaton: compute the growth function, compute the order
(if finite), detect the finiteness. The packages~\FR\ and~\SK\ provide
implementations of the three problems. Here we propose new
implementations based on a simple idea which fully uses the automaton
structure.

\subsection{Growth}\label{sec:growth}
Consider a Mealy automaton \(\aut{A}=(A,\Sigma,\delta,\rho)\) and an
element \(\mot{x}\in A^*\). The
\emph{length\/} of \(\rho_{\mot{x}}\), denoted by
\(|\rho_{\mot{x}}|\), is defined as follows: \[|\rho_{\mot{x}}| = 
\min\{ n\mid \exists \mot{y}\in A^n,\, \rho_{\mot{x}}=\rho_{\mot{y}}\}\:.\]

The \emph{growth series\/} of \(\aut{A}\) is the formal power series
given by \[\sum\limits_{g\in \presm{\aut{A}}} t^{|g|} =
\sum\limits_{n\in\N} \#\{g\in \presm{\aut{A}}\,;\, |g|=n\}\:t^n\:.\]

In words, the growth series enumerates the semigroup elements
according to their length.
This is an instanciation of the notion of spherical growth series for a finitely generated semigroup.
Observe that the series is a polynomial if and only if the semigroup
is finite.

\paragraph{Using the Generic Algorithm.}
Since the word problem is solvable, it is possible to compute an
arbitrary but finite number of coefficients of the growth series. 
Indeed for each~\(n\), generate the set of elements of
length~$n$ by multiplying elements of length~$n-1$
with generators and detecting-deleting duplicated elements by solving
the word problem. The functions~\lstinline{Growth}
from~\SK\ and~\lstinline{WordGrowth} from~\FR\ both follow this pattern.
Therefore the structure of the underlying Mealy automaton is used only
to get a solution to the word problem (in fact, both
\lstinline{Growth} and \lstinline{WordGrowth} are generic, in the
sense that they are applicable for any (semi)group with an
 implemented solution to the word problem).

\paragraph{New Implementation.}
We propose a new implementation based on a simple observation. Knowing
the elements of length~\(n-1\), Nerode minimization can be used in a
global manner to obtain simultaneously the elements of
length~\mbox{\(n\)}. Concretely, with each integer~$n\geq 1$ is
associated a new Mealy automaton~$\tensorpower{\aut{A}}{n}$ defined
recursively as follows:
\[\tensorpower{\aut{A}}{n}=\mz(\tensorpower{\aut{A}}{n-1}\times\mz(\aut{A}))
\qquad\hbox{and}\qquad
\tensorpower{\aut{A}}{1}=\mz(\aut{A})\:.\]

Here, we assume, without real loss of generality, that
the identity element is one of the generators (otherwise simply add a
new state to the Mealy automaton coding the identity). This way, the
elements of \(\aut{A}_n\) are exactly the elements of length at most~\(n\).

\begin{lstlisting}[language=GAP]
AutomatonGrowth:= function(arg)
local aut, radius, growth, sph, curr, next, r;
aut:=arg[1];   # Mealy automaton
if Length(arg)>1 then radius:=arg[2];
                 else radius:=infinity;
fi;
r := 0;    curr := TrivialMealyMachine([1]);
next := Minimized(aut);
aut := Minimized(next+TrivialMealyMachine(Alphabet(aut)));
sph := aut!.nrstates - 1;  # number of non-trivial states
growth := [next!.nrstates-sph];
while sph>0 and r<radius
do  Add(growth,sph);
    r := r+1;    curr := next;
    next := Minimized(next*aut);
    sph := next!.nrstates-curr!.nrstates;				
od;
return growth;
end;
\end{lstlisting}

Note that~\lstinline{AutomatonGrowth(aut)} computes the growth of the semigroup~$\pres{\tt aut}_+$,
while \lstinline{AutomatonGrowth(aut+aut^-1)} computes the growth of the group~$\pres{\tt aut}$.

\paragraph{Experimental Results.}
First we run~\lstinline{AutomatonGrowth} and~\FR's
\lstinline{WordGrowth} on the Grigorchuk automaton, a famous Mealy
automaton generating an infinite group. For~radius~10, \lstinline{AutomatonGrowth} is much faster, 76~ms as opposed to
\numprint{9912}~ms\footnote{All timings displayed
  in this paper
  have been obtained on an Intel Core~2~Duo computer with clock speed
  3,06~GHz.\label{fn:macMP}}. The explanation is simple: \lstinline{WordGrowth}
calls the minimization procedure \numprint{57577} times while
\lstinline{AutomatonGrowth} calls it only 12 times.
Here are the details.

\begin{lstlisting}[language=GAP]
gap> aut := GrigorchukMachine;; radius:= 10;;
gap> ProfileFunctions([Minimized]);
gap> WordGrowth(SCSemigroupNC(aut), radius); time;
[ 1, 4, 6, 12, 17, 28, 40, 68, 95, 156, 216 ]
9912
gap> DisplayProfile();
  count  self/ms  chld/ms  function                                           
  57577     7712        0  Minimized                                          
            7712           TOTAL         
gap> ProfileFunctions([Minimized]);
gap> AutomatonGrowth(aut, radius); time;
[ 1, 4, 6, 12, 17, 28, 40, 68, 95, 156, 216 ]
76
gap> DisplayProfile();
  count  self/ms  chld/ms  function                                           
     12       72        0  Minimized                                          
              72           TOTAL            
\end{lstlisting}

\medskip

Now we compare the running times of the implementations for the
computation of the first terms of the growth series for all 335
bireversible 3-letter 3-state Mealy automata (up to equivalence). In Tab.~\ref{BenchGrowthBIR33}, some computations
with~\FR's~\texttt{WordGrowth} or with~\SK's \texttt{Growth} could not
be completed in reasonable time for radius 7.

\begin{table}[h]
\centering\footnotesize
\caption{Average time (in ms)}
\setstretch{1.2}
\begin{tabular}{|c|r|r|r|r|r|r|r|r|}
	\hline
radius	&\multicolumn{1}{c|}{1}	&\multicolumn{1}{c|}{2}	&\multicolumn{1}{c|}{3}	&\multicolumn{1}{c|}{4}	&\multicolumn{1}{c|}{5}	&\multicolumn{1}{c|}{6}	&\multicolumn{1}{c|}{7}\\	
	\hline
	\FR's \texttt{WordGrowth}	
		&~3.4~ 	&~29.0~ 		&~555.0~	&~\numprint{8616}.5~	&~\numprint{131091}.4~		&~\numprint{2530170}.3~		&~~~~?~~~~~\\
	\hline
	~\SK's \texttt{Growth}~
		&0.7~	&2.8~		&16.9~	&158.9~				&\numprint{1909}.0~		&~\numprint{22952}.8~ 		&~~~~?~~~~~\\
	\hline
	\texttt{AutomatonGrowth}	
		&0.6~ 	&1.8~ 		&5.9~ 	&28.9~ 				&\numprint{187}.3~ 			&\numprint{1005}.9~		&~\numprint{7131}.4~\\
	\hline
\end{tabular}
\label{BenchGrowthBIR33}
\end{table}%

\vspace*{-.7cm}

\subsection{Order of the (Semi)group}\label{sec:order}
Although the finiteness problem is still open, some semidecision procedures
enable to find the order of an expected finite (semi)group.
\FR\ and~\SK\ use orthogonal approaches. Our new implementation
refines the one of~\FR\ and remains orthogonal to the one of~\SK.

\paragraph{\SK's Implementation.}
The \GAP\ package \SK\ 
provides the function \lstinline{LevelOfFaithfulAction}, which allows
to compute---very efficiently in some cases---the order of the
generated group. The principle is the following.
Let \(\aut{A}=(A,\Sigma,\delta,\rho)\) be
an invertible Mealy automaton and let \(G_k\) be the group generated
by the restrictions of the production functions to
\(\Sigma^k\). If \(\#G_k=\#G_{k+1}\) for some \(k\), then
\(\pres{\aut{A}}\) is finite of order~\(\#G_k\).
This function can be easily adapted to a non-invertible Mealy
automaton.

Observe that \lstinline{LevelOfFaithfulAction} cannot be used to
compute the growth series. Indeed at each step a
quotient of the (semi)group is computed. On the other hand 
\lstinline{LevelOfFaithfulAction} is a good bypass strategy for the
order computation. Furthermore, it takes advantage from the special
ability of~\GAP\ to manipulate permutation groups. 

\paragraph{\FR's Implementation and the New Implementation.}
Any algorithm computing the growth series can be used to compute the
order of the generated (semi)group if finite. It suffices to compute
the growth series until finding a coefficient equal to zero.
This is the approach followed by~\FR. Since we proposed, in the previous
section, a new implementation to compute the growth series, we obtain
as a byproduct a new procedure to compute the order. We call it
\lstinline{AutomSGrOrder}.

\paragraph{Experimental Results.}

The orthogonality of the two previous approaches can be simply illustrated by recalling the introductory example of~Fig.~\ref{monster}.
Neither \FR's \lstinline{Order} nor~\lstinline{AutomSGrOrder} are able to compute the order of the large group,
while \SK\ via~\lstinline{LevelOfFaithfulAction} succeeds in only~\numprint{14338}~ms.
Conversely, \lstinline{AutomSGrOrder} computes the order of the
small semigroup in~17~ms, while 
an adaptation of~\lstinline{LevelOfFaithfulAction} (to non-invertible
Mealy automata) takes \numprint{2193}~ms.

\subsection{Finiteness}
There exist several criteria to detect the finiteness of an automaton
(semi)group, see~\cite[...]{AKLMP12,anto,sidkiconjugacy,sidki,sst}. But the
decidability of the finiteness is still an open question.
Each procedure to compute the order of a (semi)group
yields a semidecision procedure for the finiteness
problem.
Both packages~\FR\ and~\SK\
apply a number of previously known criteria of (in)finiteness and then
intend
to conclude by ultimately using an order computation.

We propose an additional ingredient which uses minimization in
a subtle way. Here, the semigroup to be tested is successively
replaced by new ones which are finite if and only if the original one
is finite. It is possible 
to incorporate this ingredient to get two new implementations, one in
the spirit of~\FR\ and one in the spirit of~\SK.  
The new implementations are order of magnitudes better
than the old ones. Both are useful since the fastest one depends on the
cases.

\subsubsection{\(\mz\dz\)-reduction of Mealy Automata and Finiteness}
\label{s:reduction}
The \(\mz\dz\)-reduction was introduced in~\cite{AKLMP12} to give a
sufficient condition of finiteness. The new semidecision procedures start
with this reduction.

\begin{definition}
A pair of dual Mealy automata is \emph{reduced} if both automata
are minimal. Recall that $\mz$ (\resp\ \(\dz\)) is the operation of
minimization (\resp\ dualization). The \emph{$\mz\dz$-reduction} of a Mealy
automaton \(\aut{A}\) consists in minimizing the automaton or its dual
until the resulting pair of dual Mealy automata is reduced.
\end{definition}

The \(\mz\dz\)-reduction is well-defined:
if both a Mealy automaton and its dual automaton are non-minimal,
the reduction is confluent~\cite{AKLMP12}.
An example of $\mz\dz$-reduction is given in
Fig.~\ref{fig-md-reduction}.

\begin{figure}[h]
\centering
\SmallPicture
\FixVCGridScale{.7}
\VCDraw{%
\begin{VCPicture}{(-3,-19)(26,-5)}
\State[a]{(0,-8)}{AA} \State[b]{(6,-8)}{BB}
\VCPut{(8,-9)}{\huge\(\aut{A}\)}
\ArcL{AA}{BB}{\StackTwoLabels{\IOL{0}{1}}{\IOL{2}{3}}}
\ArcL{BB}{AA}{\StackTwoLabels{\IOL{0}{3}}{\IOL{2}{1}}}
\LoopN[.2]{AA}{\StackTwoLabels{\IOL{1}{0}}{\IOL{3}{2}}}
\LoopN[.8]{BB}{\StackTwoLabels{\IOL{1}{0}}{\IOL{3}{2}}}
\Point{(10,-8)}{E2} \Point{(13,-8)}{F2}
\EdgeL{E2}{F2}{{\mathfrak d}}
%
%
\State[0]{(17,-5)}{A0} \State[1]{(23,-5)}{A1}
\State[3]{(17,-9)}{A3} \State[2]{(23,-9)}{A2}
\ArcL[.3]{A1}{A0}{\StackTwoLabels{\IOL{a}{a}}{\IOL{b}{b}}}
\ArcL{A0}{A1}{\IOL{a}{b}}
\ArcL[.3]{A3}{A2}{\StackTwoLabels{\IOL{a}{a}}{\IOL{b}{b}}}
\ArcL{A2}{A3}{\IOL{a}{b}}
\EdgeR{A0}{A3}{\IOL{b}{a}}
\EdgeR{A2}{A1}{\IOL{b}{a}}
\Point{(20,-11)}{E3} \Point{(20,-13)}{F3}
\EdgeL{E3}{F3}{{\mathfrak m}}
%
%
\StateVar[13]{(17,-16)}{A13} \StateVar[02]{(23,-16)}{A02}
\ArcL{A13}{A02}{\StackTwoLabels{\IOL{a}{a}}{\IOL{b}{b}}}
\ArcL{A02}{A13}{\StackTwoLabels{\IOL{a}{b}}{\IOL{b}{a}}}
\Point{(13,-16)}{E4} \Point{(10,-16)}{F4}
\SetEdgeLineStyle{dashed}
\EdgeR{E4}{F4}{\mathfrak{dmdmd}}
\SetEdgeLineStyle{solid}
\StateVar[ab]{(3,-16)}{Z}
\VCPut{(7,-17)}{\huge\(\mdreduc{\aut{A}}\)}
\LoopN[.2]{Z}{\IOL{0123}{0123}}
\end{VCPicture}}
\caption{The $\mz\dz$-reduction of a pair of dual Mealy
  automata}\label{fig-md-reduction}
\end{figure}

The sequence of mini\-mi\-zation-dualization can be arbitrarily long: 
the minimization of a Mealy automaton with a minimal dual
can make the dual automaton non-minimal.

If \(\aut{A}\) is a Mealy automaton, we denote by
\(\mdreduc{\aut{A}}\) the corresponding Mealy automaton after
\(\mz\dz\)-reduction.

\begin{theorem}[\cite{AKLMP12}]\label{cor-md-triviality}
A Mealy automaton \(\aut{A}\) generates a finite (semi)group
if and only if $\mdreduc{\aut{A}}$ generates a finite (semi)group.
\end{theorem}

This is the starting point of the new implementations.
We use an additional fact. We can prune a Mealy automaton by deleting
the states which are not accessible from a cycle. This does not change
the finiteness or infiniteness of the generated
(semi)group~\cite{anto}.

\subsubsection{The New Implementations}\label{sec:implementations}

The design of procedure \lstinline{IsFinite1} is consistent
with the one of~\lstinline{AutomatonGrowth}. Hence
\lstinline{IsFinite1} is much closer to~\FR\ than to~\SK.
Here we propose a version that works with the automaton and its dual in parallel.

\begin{lstlisting}[basicstyle=\ttfamily\footnotesize]
IsFinite1 := function (aut, limit)
local radius, dual, curr1, next1, curr2, next2;
radius := 0;
aut := MDReduced(Prune(aut));    dual := DualMachine(aut);
curr1 := MealyMachine([[1]],[()]);    curr2 := curr1;
next1 := aut;    next2 := dual;
while curr2!.nrstates<>next2!.nrstates and radius<limit
do  radius := radius + 1;    curr1 := next1;
    next1 := Minimized(next1*aut);
    if curr1!.nrstates<>next1!.nrstates
    then  curr2 := next2;
          next2 := Minimized(next2*dual);
    else  return true;
    fi;
od;
if curr2!.nrstates = next2!.nrstates then  return true; fi;
return fail;
end;
\end{lstlisting}

The procedure \lstinline{IsFinite2} is a refinement of~\SK's
\lstinline{LevelOfFaithfulAction}: the minimization is called on the
dual and can be enhanced again to work in parallel on the Mealy
automaton and its dual.

\begin{lstlisting}[language=GAP]
IsFinite2 := function(aut,limit)
local f1, f2, next, cs, ns, lev;
aut := MDReduced(Prune(aut));
if IsInvertible(aut) then f1:=Group; f2:=PermList;
                     else f1:=Semigroup; f2:=Transformation;
fi;
lev := 0; cs := 1;	ns := Size(f1(List(aut!.output,f2)));
aut := DualMachine(aut);    next := aut;
while cs<ns and lev<limit
do lev := lev+1;   cs := ns;   next := Minimized(next*aut);
   ns := Size(f1(List(DualMachine(next)!.output,f2)));
od;
if cs=ns then return true; else return fail; fi;
end;
\end{lstlisting}

\paragraph{Experimental Results.}

Tab.~\ref{BenchFinitenessSmall} presents the average time to detect
finiteness of (semi)groups generated by $p$-letter $q$-state
invertible or reversible Mealy automata with~$p+q\in\{5,6\}$. To get a
fair comparison of the implementations, what is given is the minimum
of the running times for an automaton and its dual (see Theorem~\ref{finitedualfinite}). 

\begin{table}[h]
\centering\footnotesize
\caption{Average time (in ms) to detect finiteness of (semi)groups}
\setstretch{1.2}
\begin{tabular}{|m{7mm}|m{7mm}|m{7mm}|m{7mm}||m{7mm}|m{7mm}|m{7mm}|m{7mm}||m{7mm}|m{7mm}|m{7mm}|m{7mm}|}
\hline
\multicolumn{4}{|c||}{2-~3-} & \multicolumn{4}{c||}{2-~4-} & \multicolumn{4}{c|}{3-~3-}\\ \hline
  \multicolumn{1}{|c|}{\FR}&  \multicolumn{1}{c|} {\tt aut} &   \multicolumn{1}{c|}{\tt Fin1} &   \multicolumn{1}{c||}{\tt Fin2} & 
    \multicolumn{1}{c|}{\FR} &   \multicolumn{1}{c|}{\tt aut} &   \multicolumn{1}{c|}{\tt Fin1} & \multicolumn{1}{c||}{\tt Fin2} & 
    \multicolumn{1}{c|}{\FR} &   \multicolumn{1}{c|}{\tt aut} &   \multicolumn{1}{c|}{\tt Fin1} &  \multicolumn{1}{c|} {\tt Fin2}\\ \hline
\multicolumn{1}{|c|}{0.68} 			&\multicolumn{1}{c|} {0.81} 		&\multicolumn{1}{c|} {0.49} 	&\multicolumn{1}{c||} {0.49}
&\multicolumn{1}{c|}{36.36} 			&\multicolumn{1}{c|} {1.79}		&\multicolumn{1}{c|} {0.52} 	&\multicolumn{1}{c||} {0.62} 
&\multicolumn{1}{c|}{\numprint{1342}.12}	&\multicolumn{1}{c|} {3.78}		&\multicolumn{1}{c|} {0.61}	&\multicolumn{1}{c|} {0.70}\\ \hline
\end{tabular}
\\{\footnotesize \FR: \FR's \lstinline{IsFinite};\hfill \lstinline{aut}: \SK's \lstinline{IsFinite};\hfill \lstinline{Fin1}: \lstinline{IsFinite1};\hfill \lstinline{Fin2}: \lstinline{IsFinite2}}
\label{BenchFinitenessSmall}
\end{table}

\vspace*{-.8cm}

\section{Conjectures}\label{sec:conjectures}
The efficiency of the new implementations enables to carry out
extensive experimentations. We propose several conjectures
supported by these experiments. 

\medskip

Recall the example given in the introduction. The (semi)groups
generated by the Mealy automaton and its dual were strikingly
different, with a very large one and a rather small one. This
seems to be a general fact that we can state as an informal
conjecture:

 \emph{Whenever a Mealy automaton generates a finite
  (semi)group which is very large with respect to the number of states
  and letters of the
  automaton, then its dual  generates a small one. }

\smallskip

\emph{Observation:} Any pair of finite (semi)groups can be generated
by a pair of dual Mealy automata, see~\cite[Prop.~9]{AKLMP12}. The
standard construction leads to automata whose sizes are related to
the orders of the (semi)groups. Therefore it does not contradict the
informal conjecture.

\smallskip

\input{equilibrium33.tex}

Fig.~\ref{fig:equilibrium33} illustrates this informal conjecture:
for \(\aut{A}\) covering the set of all 3-letter 3-state invertible
Mealy automata, the
endpoints of each segment represent respectively the order of
\(\presm{\aut{A}}\) and of \(\presm{\dual{\aut{A}}}\), for all pairs
detected as being finite.

To assess finiteness, the procedures~\lstinline{IsFinite1}
and~\lstinline{IsFinite2} have been used. If
the tested Mealy automaton and its dual were both found to have
more than 4000~elements,
the procedures were stopped, and the (semi)groups were
supposed to be infinite. 
Based on the informal conjecture, we believe to have captured
all finite groups. If true:

\smallskip

\begin{itemize}
\item There are~\numprint{14089} Mealy automata generating finite (semi)groups
among the \numprint{233339} invertible or reversible 3-letter 3-state Mealy automata;
\item The group generated by Fig.~\ref{monster}-right is the largest finite group.
\end{itemize}

\begin{figure}[t]
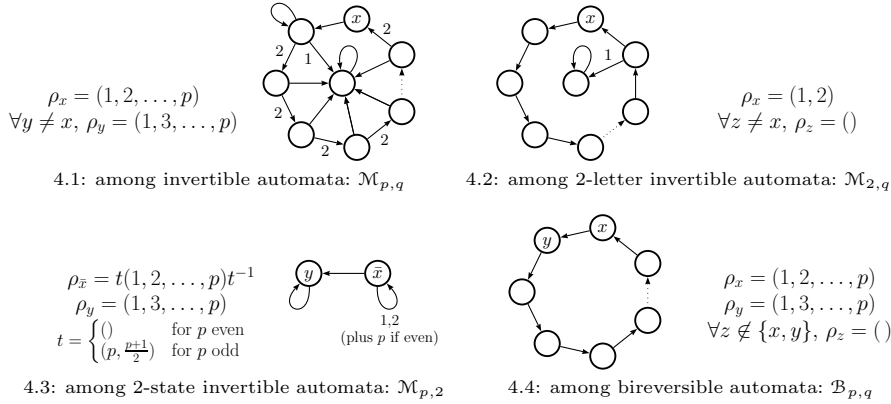

\centering
\subfigure[\label{Mpq}among invertible automata: \(\aut{M}_{p,q}\)]{
\SmallPicture
\FixVCScale{.4}
\FixVCGridScale{2.2}
\VCDraw{\begin{VCPicture}{(-4.5,-1.3)(1.5,.5)}
\State{(-.62,.78)}{A1} \State[x]{(.22,.97)}{AQ}
\State{(-1,0)}{A2}  \State{(0,0)}{A0} \State{(.9,-.43)}{A5}\State{(.9,.43)}{A6}
\State{(-.62,-.78)}{A3} \State{(.22,-.97)}{A4} 
\EdgeR{A1}{A2}{$\footnotesize2$}
\EdgeR{A2}{A3}{$\footnotesize2$}
\EdgeR{A3}{A4}{$\footnotesize2$}
\EdgeR{A4}{A5}{$\footnotesize2$}
\EdgeR{A6}{AQ}{$\footnotesize2$}
\EdgeR[.3]{A1}{A0}{$\footnotesize1$}
	\CLoopR{135}{A1}{}
	\CLoopR{75}{A0}{}
	\EdgeR{AQ}{A1}{}
	\EdgeR{A2}{A0}{}
	\EdgeR{A3}{A0}{}
	\EdgeR{A4}{A0}{}
	\EdgeR{A5}{A0}{}
	\EdgeR{A4}{A0}{}
	\EdgeR{A5}{A0}{}
	\EdgeR{A6}{A0}{}
\SetEdgeLineStyle{dotted}
	\EdgeR{A5}{A6}{}
\RstEdgeLineStyle
\VCPut{(-3.3,-.2)}{\huge \(\rho_x = (1,2,\dots,p)\)}
\VCPut{(-3.3,-.6)}{\huge \(\forall y\neq x,\,\rho_y = (1,3,\dots,p)\)}
\end{VCPicture}}
}
\subfigure[\label{M2q}among 2-letter invertible automata: \(\aut{M}_{2,q}\)]{
\SmallPicture
\FixVCScale{.4}
\FixVCGridScale{2.2}
\VCDraw{\begin{VCPicture}{(-1.5,-1.3)(5,.5)}
\State{(-.62,.78)}{A1} \State[x]{(.22,.97)}{AQ}
\State{(-1,0)}{A2}  \State{(0,0)}{A0} \State{(.9,-.43)}{A5}\State{(.9,.43)}{A6}
\State{(-.62,-.78)}{A3} \State{(.22,-.97)}{A4} 
\EdgeR{A1}{A2}{}
\EdgeR{A2}{A3}{}
\EdgeR{A3}{A4}{}
\EdgeR{A5}{A6}{}
\EdgeR{A6}{AQ}{}
\EdgeR[.3]{A6}{A0}{$\footnotesize1$}
	\CLoopR{75}{A0}{}
	\EdgeR{AQ}{A1}{}
\SetEdgeLineStyle{dotted}
	\EdgeR{A4}{A5}{}
\RstEdgeLineStyle
\VCPut{(3.2,-.2)}{\huge \(\rho_x = (1,2)\)}
\VCPut{(3.2,-.6)}{\huge \(\forall z\neq x,\,\rho_z = ()\)}
\end{VCPicture}}
}
\subfigure[\label{Mp2}among 2-state invertible automata: \(\aut{M}_{p,2}\)]{
\SmallPicture
\FixVCScale{.4}
\FixVCGridScale{2.2}
\VCDraw{\begin{VCPicture}{(-5,-1.3)(1.5,1.1)}
\AlignedLabel
\State[y]{(-0.82,.28)}{A1} \State[\bar x]{(.22,.28)}{AQ}
\EdgeR{AQ}{A1}{}
	\CLoopL{250}{A1}{}
	\CLoopR[.4]{290}{AQ}{}
	\VCPut{(.4,-.45)}{\Large 1,2}
	\VCPut{(.4,-.7)}{\Large\hbox{(plus $p$ if even)}}
	\VCPut{(-3.05,.2)}{\huge \(\rho_{\bar x} = t(1,2,\dots,p)t^{-1}\)}
	\VCPut{(-3.2,-.2)}{\huge \(\rho_{y} = (1,3,\dots,p)\)}
	\VCPut{(-3.2,-.75)}{\setstretch{.2}\LARGE \(t =
					\begin{cases}	()				&\hbox{for~$p$ even}\\
								(p,\frac{p+1}{2})	&\hbox{for~$p$ odd}\end{cases}\)}
\end{VCPicture}}
}\subfigure[\label{Bpq}among bireversible automata: \(\aut{B}_{p,q}\)]{
\SmallPicture
\FixVCScale{.4}
\FixVCGridScale{2.2}
\VCDraw{\begin{VCPicture}{(-1.5,-1.3)(5,1.1)}
\State[x]{(.22,.97)}{AQ}
\AlignedLabel
\State[y]{(-.62,.78)}{A1} 
\State{(-1,0)}{A2}  \State{(.9,-.43)}{A5}\State{(.9,.43)}{A6}
\State{(-.62,-.78)}{A3} \State{(.22,-.97)}{A4} 
\EdgeR[.7]{A1}{A2}{}
\EdgeR[.3]{A2}{A3}{}
\EdgeR{A3}{A4}{}
\EdgeR{A4}{A5}{}
\EdgeR{A6}{AQ}{}
\EdgeR{AQ}{A1}{}
\SetEdgeLineStyle{dotted}
\EdgeR{A5}{A6}{}
\RstEdgeLineWidth
\VCPut{(3.2,.2)}{\huge \(\rho_{x} = (1,2,\dots,p)\)}
\VCPut{(3.2,-.2)}{\huge \(\rho_{y} = (1,3,\dots,p)\)}
\VCPut{(3.2,-.6)}{\huge \(\forall z\not\in\{x,y\},\,\rho_z = (\,)\)}
\end{VCPicture}}
}\label{monster-family}
\caption{Automata conjectured to generate the largest finite automaton groups}
\end{figure}

\medskip Our next conjectures are concerned with the largest finite groups that
can be generated by automata of a given size.

Consider the family of \(p\)-letter \(q\)-state Mealy automata~\((\aut{M}_{p,q})_{p+q>5}\)
displayed on Fig.~\ref{Mpq} for~\(p>2\) and~\(q>2\), while the specializations for~\(p=2\) and~\(q=2\) are displayed on Fig.~\ref{M2q} and Fig.~\ref{Mp2}.
The example of Fig.~\ref{monster}-right is
\(\aut{M}_{3,3}\).

\begin{conjecture}\label{conj:inv}
The group \(\pres{\aut{M}_{p,q}}\) is finite. Every \(p\)-letter
$q$-state invertible Mealy automaton generates a group which is either
infinite or has an order smaller than~\(\#\pres{\aut{M}_{p,q}}\). 
\end{conjecture}

If true, Conjecture~\ref{conj:inv} implies the decidability of the
finiteness problem for automaton groups. Without entering
into the details of the experimentations, we
consider that Conj.~\ref{conj:inv} is reasonably well supported
for~\(p+q<9\). As for actually computing 
\(\#\pres{\aut{M}_{p,q}}\), here are the only cases with~\(q>2\) for which we
succeeded: 
\begin{align*}
\forall q,\, 4\leq q\leq 8,\qquad
\#\pres{\aut{M}_{2,q}}=2^{2^{q-1}+\frac{(q-2)(q-1)}{2}-2}\:,\qquad
\qquad \\
\#\pres{\aut{M}_{3,3}} = 2^{64}\cdot 3^{4}, \qquad \#\pres{\aut{M}_{3,4}} = 2^{325}\cdot 3^{13},\qquad\#\pres{\aut{M}_{4,3}} = 2^{288}\cdot 3^{422}\:.
\end{align*}
These groups are indeed huge. Incidentally, the finiteness of~\(\pres{\aut{M}_{p,q}}\) is checked for~\(p+q<11\) and the informal
conjecture is supported further by computing
the order of  the much smaller semigroups generated by the duals:
\begin{center}
\footnotesize
\begin{tabular}{|c|c|c|c|c|c|c|c|}
\hline
$\#\pres{\dual{\aut{M}_{p,q}}}_+$  &\gray ~~2~~& 3 & 4 & 5 & 6 & 7 & 8\tabularnewline \hline
\gray 2	&\gray- 	&\gray- 	&\gray219 			&\gray\numprint{1759} 	&\gray \numprint{13135} &\gray\numprint{94143}	&\gray\numprint{656831}\tabularnewline
3	&\gray- 	& 238	& \numprint{1552} 	& \numprint{8140} 	& \numprint{37786} & \numprint{162202}	& $\cdots$	\tabularnewline
4	&\gray89 		&\numprint{1381}	&\numprint{12309}	& \numprint{87125}	& \numprint{543061}& $\cdots$& $\cdots$\tabularnewline
5	&\gray131 	&\numprint{6056}	&\numprint{67906}	& \numprint{602656}& $\cdots$& $\cdots$& $\cdots$\tabularnewline
6	&\gray337	&\numprint{22399}	&\numprint{302011}	& $\cdots$& $\cdots$& $\cdots$& $\cdots$\tabularnewline
7	&\gray351	&\numprint{74194}	& $\cdots$& $\cdots$& $\cdots$& $\cdots$& $\cdots$\\
\hline
\end{tabular}
\end{center}

\medskip

Experimentally, the finite groups generated by
bireversible Mealy automata seem to be much smaller.
Consider the family of bireversible
automata~\((\aut{B}_{p,q})_{p,q}\) of
Fig.~\ref{Bpq}. The group~\(\pres{\aut{B}_{p,q}}\) is
isomorphic to~\({\mathfrak S}_p^q\), while the
group~\(\pres{\dual{\aut{B}_{p,q}}}\) is isomorphic
to~$\mathbb{Z}_q$. Again, the following is reasonably well supported for~\(p+q<9\):

\begin{conjecture}\label{conj:bir}
Every \(p\)-letter \(q\)-state bireversible Mealy automaton generates a group which is either
infinite or has an order smaller than~\(\#\pres{\aut{B}_{p,q}}=p!^q\). 
\end{conjecture}

Our last conjecture is of a different nature and deals with the
structure of infinite automaton semigroups.

\begin{conjecture}\label{conj:2state}
Every $2$-state reversible Mealy automaton generates a semigroup
which is either finite or free of rank~2. 
\end{conjecture}

The conjecture has been tested and seems correct for reversible 2-state Mealy 
automata up to 6 letters.
In the experiments, a semigroup generated by a $p$-letter automaton is conjectured
to be free if its growth series coincides with~$(2t)^n$ up to
radius~$p^2/2$ and if its dual generates a seemingly infinite
group.  

\bibliographystyle{plain}
\bibliography{./finiteness}

\end{document}